\documentclass[preprint,superscriptaddress, showkeys, showpacs ] {revtex4}

\begin{document}
\title{Secure Direct Communication Based on Secret Transmitting Order of Particles}
\author{Ai-Dong Zhu}
\affiliation{Center for the Condensed-Matter Science and Technology,
Harbin Institute of Technology, Harbin, Heilongjiang  150001,
People's Republic of China} \affiliation{Department of Physics,
College of Science, Yanbian University, Yanji, Jilin 133002,
People's Republic of China}
\author{Yan Xia}
\author{Qiu-Bo Fan}
\affiliation{Department of Physics, College of Science, Yanbian
University, Yanji, Jilin 133002, People's Republic of China}
\author{Shou Zhang\footnote{E-mail: szhang@ybu.edu.cn}}

\affiliation{Center for the Condensed-Matter Science and Technology,
Harbin Institute of Technology, Harbin, Heilongjiang  150001,
People's Republic of China}

\affiliation{Department of Physics, College of Science, Yanbian
University, Yanji, Jilin 133002, People's Republic of China}

\begin{abstract}
We propose the schemes of quantum secure direct communication
(QSDC) based on secret transmitting order of particles. In these
protocols, the secret transmitting order of particles ensures the
security of communication, and no secret messages are leaked even
if the communication is interrupted for security. This strategy of
security for communication is also generalized to quantum
dialogue. It not only ensures the unconditional security but also
improves the efficiency of communication.
\end{abstract}

\maketitle

\section{Introduction}
With the rapid development of information technology, quantum
cryptography has been an important and attractive study area. It
ensures that the secret message is intelligible only for the two
legitimated parties of communication without being altered or
stolen. Since Bennett and Brassard proposed BB84 protocol
\cite{bb84} which is a proven secure protocol, many quantum key
distribution (QKD) schemes have been proposed and the experimental
feasibility of them is also discussed
\cite{Bennett,Gold,Koashi,Hwang,Cabello1,Cabello2,Xue,Song}.
Although the methods used in these schemes are various, the basic
principle is the same, i.e., the two remote legitimated users
(Alice and Bob) establish a shared secret key through the
transmission of quantum signals, after this they can use this key
to encrypt or decrypt the secret messages. This means the two
parties have to share a secret key before the secret message is
transmitted. As for communication, this beforehand step
undoubtedly reduces the efficiency of communication. Our motive to
build quantum channel is not only to transmit information securely
without being eavesdropped on but also to improve the efficiency
of communication.

In recent years, a novel scheme, quantum secure direct communication
(QSDC) has been proposed and pursued
\cite{Beige,Bostrom,cai,Deng04,Deng03,Yan}. In this scheme the
transmitted message can be read only after a final transmission of
an additional classical information without first establishing a
shared secret key. In 2002, Bostr\"{o}m and Felbinger \cite{Bostrom}
proposed a Ping-Pong QSDC protocol using EPR pairs as quantum
message carriers, which is insecure for a noisy quantum channel as
shown by W\'{o}jcik \cite{wojcik}. Cai and Deng gave a scheme using
single photons as a quantum one-time pad to encode the secret
messages \cite{cai,Deng04}. Meanwhile, Deng et al put forward a
two-step QSDC protocol using blocks of EPR pairs \cite{Deng03}. In
this two-step scheme the EPR pairs are divided into two sequences,
checking-sequence and message-sequence, which are sent by two steps,
and the receiver need to check the security of the channel twice
(one for checking-sequence and another for message-sequence). In
this paper, two QSDC schemes based on transmitting order of
particles are proposed. In these two schemes, we also use EPR pairs
as the messages carriers, but the transmitting order of particles is
secret to any other people except the sender himself(herself), so
the eavesdropper (Eve) is not able to get any secret messages by
performing a valid measurement. And we need checking security only
once. Furthermore, we also apply this strategy of secret
transmitting order to bidirectional communication, which is
so-called quantum dialogue \cite{NBA04PLA}. Our present schemes not
only ensure the unconditional security but also improve the
efficiency of communication. The concrete protocols for QSDC are
given in Section 2. In Section 3 the security of the strategy is
discussed. In Section 4 we generalize the application of the
strategy based on secret transmitting order to quantum dialogue.
Finally, we give a discussion and summary on the present schemes.
\section{Schemes for QSDC}
 An EPR pair can be in one of the following four states,
\begin{equation}
|\Psi^{\pm}\rangle=\frac{1}{\sqrt{2}}(|0\rangle_{i}|1\rangle_{i^{'}}\pm|1\rangle_{i}|0\rangle_{i^{'}}),
\end{equation}
\begin{equation}
|\Phi^{\pm}\rangle=\frac{1}{\sqrt{2}}(|0\rangle_{i}|0\rangle_{i^{'}}\pm|1\rangle_{i}|1\rangle_{i^{'}}),
\end{equation}
where $|0\rangle$ and $|1\rangle$ are eigenvectors of Pauli
operator $\sigma_{z}$. The subscripts $i$ and $i'$ stand for the
two correlated particles of an EPR pair. Firstly, Alice and Bob
agree on that the four local operations $U_0=I=|0\rangle\langle
0|+|1\rangle\langle 1|$, $U_1=\sigma_{z}=|0\rangle\langle
0|-|1\rangle\langle 1|$, $U_2=\sigma_{x}=|0\rangle\langle
1|+|1\rangle\langle 0|$ and $U_3=i\sigma_{y}=|0\rangle\langle
1|-|1\rangle\langle 0|$ represent two bits classical information
00, 11, 01, and 10, respectively. Alice prepares ordered $N$ EPR
photon pairs in the same state. Here we assume this state is
$|\Psi\rangle=\frac{1}{\sqrt{2}}(|0\rangle_{i}|1\rangle_{i'}-|1\rangle_{i}|0\rangle_{i'})$.

On these preconditions, we give the following two schemes for
QSDC.

\subsection{A round trip scheme based on the secret transmitting order of particles}
(S1) Alice divides the EPR pairs into two partner-photon sequences
$[H_{1},\ H_{2}, \cdots, H_{i}, \cdots, H_{N}]$ and $[T_{1'},\
T_{2'}, \cdots, T_{i'}, \cdots, T_{N'}]$, where $H_{i}$ and $T_{i'}$
are the two photons correlated with each other in the $i$-th $(i=1,\
2,\ \cdots\ N)$ photon pair, and $H(T)$ stands for ``home (travel)".
Then she sends the T sequence to Bob.

(S2) Bob chooses a sufficiently large subset of photons randomly
in T sequence as checking set (C-set) and the rest as message set
(M-set). By performing the four unitary operations $U_i, (i=0, 1,
2, 3)$, he encodes his checking message on C-set and secret
messages on M-set, respectively.

(S3) Bob disturbs the initial order of the $T$ sequence and
returns them to Alice, that is, the rearranged order of T sequence
is completely secret to any other people but Bob himself.

(S4) After verifying Alice has received all $T$ photons, Bob
announces C-set and the secret order in it. According to these
information and the initial states, Alice can perform Bell
measurement and deduce the probable operations performed by Bob.
Then she announces her results about Bob's operations.

(S5) By comparing his checking messages with Alice's results, Bob
can decide whether a Eve is online. If Eve is online, Bob
terminates the communication. Otherwise, he exposes the secret
transmitted order of M-set according to which Alice can read the
secret messages by Bell measurement.

In this protocol, one particle (T photon) of each EPR pair
undergoes a round trip to transfer information. This makes it
impossible for Eve to get two particles of an EPR pair
simultaneously. By disturbing the original order of particles, the
security of communication is protected from the
intercept-and-resend attack. However, the efficiency of a round
trip is lower than a one-way trip after all, so we ameliorate this
protocol to the one-way protocol based on the strategy of secret
transmitting order.

\subsection{An one way scheme based on the secret transmitting order of particles}
(S1$^{'}$) After preparing EPR pairs, Alice chooses a sufficiently
large subset randomly as the checking set (C-set) and the rest
pairs as message set (M-set). Different from the above scheme,
here the C(M)-set is composed of EPR pairs but not single photons.
Then Alice encodes her secret messages on M-set and checking
messages on C-set, respectively, by performing the four operations
on one particle (e.g. the first one) of each EPR pair. For
convenience of describing, we denote the $N$ EPR pairs with
$P_{1}(1,1'), P_{2}(2,2'), \cdots,P_{i}(i, i'), \cdots,
P_{N}(N,N')$. Taking C-set for example. Assuming Alice's checking
message is $(0100101101\cdots)$, and she chooses the first $50$
EPR pairs as C-set. Then she encodes $01$ on $P_{1}(1,1')$, $00$
on $P_{2}(2,2')$, $10$ on $P_{3}(3,3')$, $\cdots$, and so on.

(S2$^{'}$) With an order known only by herself, Alice sends these
particles to Bob one by one, namely, the particles are sent as
single form but not as pairs. For instance, Alice sends the
particles with an order $S_1(2)$, $S_2(1)$, $S_3(51)$, $S_4(5')$,
$S_5(2')$, $S_6(60)$, $S_7(10)$, $S_8(1')$,$\cdots$,
$S_j(x)$,$\cdots$, $S_k(x')$,$\cdots$,$S_{2N}(y)$, where $S_j
(i)(j\in{1,2,\cdots,2N, i\in{1, 2, \cdots, N}})$ denotes Alice
sends particle $i$ with the $jth$ turn.

(S3$^{'}$) After verifying Bob has received all the $2N$
particles, Alice declares the initial state and the matching
information of two particles in C-set through a public channel.
For instance, ${S_2\sim S_8, S_1\sim S_5, \cdots, S_j\sim
S_k,\cdots}$.

(S4$^{'}$) Alice and Bob check the security of the channel. Bob
performs a Bell-basis measurement according to the information
from Alice. Comparing with the initial state, he obtains the
result messages. Then he tells Alice about his result messages
through a classical channel. By comparing Bob's result messages
with the checking messages as well as analyzing the error rate,
Alice can judge out whether Eve is on line.

(S5$^{'}$) If the channel is secure, Alice exposes the matching
information of two particles in M-set through a classical channel.
Otherwise, Alice terminates this communication and starts next one
from beginning.

(S6$^{'}$) By performing a Bell-basis measurement, Bob obtains the
secret messages.

In this protocol all the particles undergo only a one-way trip,
which greatly reduces the opportunity of the particles being
intercepted than the round trip and two-step protocol
\cite{Deng03}, and thus improves the efficiency of communication.

\section{Security of the QSDC schemes based on secret transmitting order of particles}
Firstly, the security of our present schemes are based on the
secret order of the particles, while the security of two-step
scheme \cite{Deng03} lies in the security of the transmission of
C-sequence. In a noisy channel, Eve can hide her eavesdropping in
the noise. If Alice and Bob could not detect the eavesdropper in
the transmission of C-sequence, Eve would capture easily the two
particles in each EPR pair and take Bell-basis measurement on
them, i.e. the secret messages would be leaked partly or all.
However, this situation can be avoided in our present schemes.

Eve can not only takes intercept-and-resend attack but also takes
entangle-and-measure attack in the whole communication process. In
the round trip scheme, under the condition that Eve uses a
intercept-and-resend attack, he also creates $N$ EPR pairs which
are in the same state $|\Psi\rangle_{ht}$ with $(ht)$ are Eve's
two particles correlated mutually. When Alice sends the $T$ photon
sequence to Bob, Eve intercepts these $T$ photons and sends her
$t$ photons to Bob. Bob would take $t$ for $T$ and encodes he
secret message and checking messages by performing the unitary
operations as described above. If Bob returns them to Alice with
the initial order, Eve can intercepts the $``T(t)"$ photons again
and takes Bell-basis measurement on her $ht$ pairs to learn Bob's
secret messages and checking messages. Eve applies the same
unitary operations on the $T$ photons which she intercepted and
sends them to Alice. As a result, Eve not only gain the secret
messages, but also will not be detected. However, in our scheme in
that the initial order of $T(t)$ sequence is disturbed by Bob in
the returning process, so Eve is not able to distinguish the $t$
photon corresponding to her $h$ photon. A blind encoding on $T$
sequence can be detected. In the one-way scheme, all particles are
transmitted with a secret order. Even if Eve intercepts all the
particles, it is difficult for her to distinguish the partners of
each pair and take a valid measurement. So her interception is not
useful. Particularly, it should be noticed that in the one way
scheme, only one transmission process is used. This not only
greatly reduces the opportunity of the particles to be intercepted
but also improves the efficiency of communication.

On the basis of the above analysis, our present QSDC schemes using
the strategy of secret transmitting order are secure.

\section{Generalization to quantum dialogue based on secret transmitting order}
The above protocols are mono-directional communication. Using this
strategy, we also can generalize the above QSDC schemes to a
bidirectional communication, the so-called quantum dialogue
\cite{NBA04PLA}.

Suppose that Alice and Bob has respective secret messages
consisting of $2N$ bits to transmit to the other side. They can do
according to the following steps.

 (S1) To securely carry out a secret dialogue, Alice firstly prepares a large enough number(M) of
 ordered EPR pairs, all in the same state (e.g.,
 $|\psi_{0,0}\rangle_{ht}$). Then she encodes her secret messages $M^{A}_m$ on
  $N$ particles {\it t} $(t={\it t_1},{\it t_2},{\it t_3} \cdots {\it
 t_N})$ (M-set) and the checking messages $M^{A}_c$ on the rest $(M-N)$ $t$ particles (C-set) by means of the four unitary operations above.

 (S2) Alice sends the particles string {\it ${t}$} to Bob in order. In accordance with the order of the travel
particles {\it t}, Alice stores the remaining particles {\it h}
with him.

(S3) Confirming Bob has received the sequence $t$, Alice tell Bob
the M-set and C-set. Then Bob also encodes his secret messages
$M^{B}_m$ and checking messages $M^{B}_c$ on M-set and C-set,
respectively. Then Bob disturbs the order of $t$ sequence and
returns them to Alice.

(S4) After confirming the receiving of Alice, Bob announces the
secret order of the particles $t$ of C set. Alice performs
Bell-basis measurement on particles $t$ and corresponding partners
in C-set and announces the results $R_c$. Then both Alice and Bob
can deduce the probable checking messages $m^{B}_c$ and $m^{A}_c$
of the other sides by $m^{B}_c=R_c-M^{A}_c$ and
$m^{A}_c=R_c-M^{B}_c$, respectively.

(S5) Alice and Bob publicly announce their respective true
checking messages $M^{A}_c$ and $M^{B}_c$. If the error rates of
$m^{A}_c$ versus $M^{A}_c$, $m^{B}_c$ versus $M^{B}_c$ are
relatively high, the communication should be terminated.
Otherwise, Bob announces the secret order of M-set. Then Alice
measures on the corresponding EPR pairs and publicly announces the
results $R_m$.

(S6) Alice and Bob decode the secret messages of the other side in
terms of $M^{B}_m=R_m-M^{A}_m$ and $M^{A}_m=R_m-M^{B}_m$,
respectively.

Similarly, because the transmitting order of particles is secret
before the security checking, Eve cannot perform a valid
measurement, so the unconditional security is ensured.

\section{Discussion and Summary}
The present schemes for mono-directional and bidirectional
communication are secure in an ideal lossless channel as the
analysis above. In a practical quantum channel, there are noise
and loss which will threaten the security of quantum
communication. We need to illustrate that our scheme is still
secure in a weak noisy channel. The security-checking is based on
the statistical analysis for the error rate. Under a condition of
weak noise, a higher error rate may indicates the eavesdropping.
Hence our scheme is still valid.

In the meantime, we also notice that in most protocols on direct
communication, once Eve is detected, as the communication is
terminated, the secret messages are discarded. It is noticeable
that direct communication is different from QKD. QKD allows the
secret key between the two legitimate users to be produced over
again for security. But the motive of QSDC is communicating
messages directly, discarding messages means leakage of secret. As
for our present schemes, the secret messages are hidden in the
disordered transmitting order of particles. Eve cannot get any
useful message without a correct order even if she captures the
particles. Hence, no message is leaked except the communication is
terminated, and the secret messages can be used repeatedly between
the two legitimate users.

In summary, basing on the strategy of secret transmitting order,
two novel QSDC schemes and a quantum dialogue scheme
 have been proposed. This strategy ensures the security of communication
 not only in an ideal lossless channel but also in a weak
noisy channel. Moreover, because the secret messages is impossibly
leaked even if when communication is terminated for security, the
secret messages can be transmitted repeatedly between the sender
and the receiver.


\begin{thebibliography}{999}
\bibitem{bb84} C. H. Bennett and G. Brassard, {\it Proc. IEEE Int.
Conf. on Computers, Systems and Signal Processing, Bangalore,
India} (IEEE, New York, 1984), pp. 175-179.
\bibitem{Bennett}C. H. Bennett, G. Brassard and N. D. Mermin,
Phys. Rev. Lett. {\bf 68}, 557 (1992).
\bibitem{Gold}L. Goldenberg and L. Vaidman, Phys. Rev. Lett. {\bf 75}, 1239 (1995).
\bibitem{Koashi} M. Koashi and N. Imoto, Phys. Rev. Lett. {\bf 79}, 2383 (1997).
\bibitem{Hwang}W. Y. Hwang, I. G. Koh and Y. D. Han, Phys. Lett. A
{\bf 244}, 489 (1998).
\bibitem{Cabello1}A. Cabello, Phys. Rev. A {\bf 61}, 052312
(2000).
\bibitem{Cabello2}A. Cabello, Phys. Rev. Lett. {\bf 85}, 5635
(2000).
\bibitem{Xue}P. Xue, C. F. Li and G. C. Guo, Phys. Rev. A {\bf 65},
022317 (2002).
\bibitem{Song}D. Song, Phys. Rev. A {\bf 69}, 034301
(2004).
\bibitem{Beige} A. Beige, B.-G. Englert, Ch. Kurtsiefer, and H.
Weinfurter, Acta Phys. Pol. A {\bf 101}, 357 (2002).
\bibitem{Bostrom} K. Bostr\"{o}em and T. Felbinger, Phys. Rev.
Lett. {\bf 89}, 187902 (2002).
\bibitem{cai}Q. Y. Cai and B. W. Li, Phys. Rev. A {\bf 69}, 054301
(2004).
\bibitem{Deng04} F. G. Deng and G. L. Long, Phys. Rev. A {\bf 69},
052319 (2004).
\bibitem{Deng03}F. G. Deng, G. L. Long, and X. S. Liu, Phys. Rev.
A {\bf 68}, 042317 (2003).
\bibitem{Yan}F. L. Yan and X. Q. Zhang, Eur. Phy. J. B {\bf 41},
75 (2004).
\bibitem{wojcik}A. W\'{o}jcik, Phys. Rev. Lett. {\bf 90}, 157901
(2003).
\bibitem{NBA04PLA}B. A. Nguyen, Phys. Lett. A {\bf 328}, 6(2004).
\end{thebibliography}
\end{document}